# Investigating the potentialities of Monte Carlo simulation for assessing soil water content via proximal gamma-ray spectroscopy

Marica Baldoncini[a,b,*], Matteo Albéri[a,b], Carlo Bottardi[b,c], Enrico Chiarelli[a,b], Kassandra Giulia Cristina Raptis[a,b], Virginia Strati[b,c] and Fabio Mantovani[b,c]

[a]*INFN, Legnaro National Laboratories, Viale dell'Università 2, 35020, Legnaro, Padua, Italy;*
[b]*Department of Physics and Earth Sciences, University of Ferrara, Via Saragat 1, 44121, Ferrara, Italy;*
[c]*INFN, Ferrara Section, Via Saragat 1, 44121, Ferrara, Italy*

[*]Corresponding author.
E-mail address: baldoncini@fe.infn.it (M. Baldoncini)

**Abstract**

Proximal gamma-ray spectroscopy recently emerged as a promising technique for non-stop monitoring of soil water content with possible applications in the field of precision farming. The potentialities of the method are investigated by means of Monte Carlo simulations applied to the reconstruction of gamma-ray spectra collected by a NaI scintillation detector permanently installed at an agricultural experimental site. A two steps simulation strategy based on a geometrical translational invariance is developed. The strengths of this approach are the reduction of computational time with respect to a direct source-detector simulation, the reconstruction of $^{40}$K, $^{232}$Th and $^{238}$U fundamental spectra, the customization in relation to different experimental scenarios and the investigation of effects due to individual variables for sensitivity studies. The reliability of the simulation is effectively validated against an experimental measurement with known soil water content and radionuclides abundances. The relation between soil water content and gamma signal is theoretically derived and applied to a Monte Carlo synthetic calibration performed with the specific soil composition of the experimental site. Ready to use general formulae and simulated coefficients for the estimation of soil water content are also provided adopting standard soil compositions. Linear regressions between input and output soil water contents, inferred from simulated $^{40}$K and $^{208}$Tl gamma signals, provide excellent results demonstrating the capability of the proposed method in estimating soil water content with an average uncertainty < 1%.

**Abbreviations**

Photon Field Building (PFB); Gamma Spectrum Reconstruction (GSR); Photon Field Layer (PFL); Full Spectrum Analysis - Non Negative Least Squares (FSA-NNLS).

## 1. Introduction

Starting from its primary applications to mineral exploration and geological prospecting, gamma-ray spectrometry entered the field of applied geoscience as a highly effective technique for retrieving, at different spatial resolutions, geochemical information on the basis of the distribution of radionuclides in the environment. Although early developments focused on mapping gamma radiation emitted from terrestrial radioisotopes (i.e. $^{40}$K and daughter products of the $^{238}$U and $^{232}$Th decay chains) for the identification of rare earth ores or other mineral commodities (Bristow, 1983; Killeen, 1963; Mero, 1960; Ward, 1981), progressively the exceptional capabilities of radiometric measurements in estimating soil properties have been demonstrated (Beamish, 2015; Mahmood et al., 2013; Wilford et al., 1997; Wilford and Minty, 2006). In particular, promising applications regard soil texture (Heggemann et al., 2017; Viscarra Rossel et al., 2007), clay content (Coulouma et al., 2016; Priori et al., 2013; Van der Klooster et al., 2011), cadmium contamination (Söderström and Eriksson, 2013), pH, organic carbon and plant available potassium (Dierke and Werban, 2013; Pracilio et al., 2006).

In the panorama of environmental variables affecting radiometric measurements, water content and bulk density are the most crucial factors. As water has 1.11 times as many electrons per gram compared to most soils, water is 1.11 times as effective in attenuating gamma-radiation compared to typical soils (Grasty, 1997). The expected high sensitivity of gamma spectroscopy to soil water content has triggered numerous studies which addressed a broad range of applications including soil classification (Beamish, 2013, 2014), radon flux mapping (Manohar et al., 2013; Szegvary et al., 2007) and snow water equivalent assessment (Carroll and Carroll, 1989; Peck and Bissell, 1973). Nevertheless, the potentialities of the method for monitoring soil moisture dynamics have been not fully explored yet (Bogena et al., 2015; Pereira, 2011), especially in the field of proximal sensing, which is foreseen to be an efficient strategy for filling the existing gap between punctual measurements, generally provided by in situ electromagnetic sensors (Walker et al., 2004), and remote measurements, typically performed by satellites (Brocca et al., 2017; Zeng et al., 2016).

Although in the last decades proximal gamma-ray spectroscopy experienced a boost in terms of technological and spectral analysis developments, current radiometric data processing concerning the specific topic of soil moisture assessment is typically based on first order analytical models (Carroll, 1981; Grasty, 1997; Loijens, 1980). These methods lack however a custom approach able to integrate individual site characteristics to distinct experimental set up features. In this perspective, Monte Carlo simulations can overcome the limits of analytical solutions, which generally address the description of the sole unscattered gamma-ray flux, by providing information on the entire gamma spectra (Allyson and Sanderson, 1998; Androulakaki et al., 2016; Vlastou et al., 2006). In a Monte Carlo simulation all parameters can be separately controlled and uncertainties coming from temporary variations in the experimental conditions can be excluded, which is particularly relevant in relation to calibration procedures and feasibility studies (Chirosca et al., 2013; De Groot et al., 2009; Van der Graaf et al., 2011). This peculiarity makes the methodology highly versatile in terms of input boundary conditions and extraordinarily effective in both investigating the effects of individual variables (e.g. for sensitivity studies) and in the calibration of different source-detector systems (e.g. permanent stations, carborne based platforms).

The focus of this paper is investigating by means of Monte Carlo simulations the potentialities of proximal gamma-ray spectroscopy applied to the estimation of soil water content in precision agriculture. After providing depth and lateral horizons of proximal gamma-ray spectroscopy in section 2, we present in section 3 a strategy which allows to tackle the challenge of simulating gamma spectra generated by a homogeneous infinite medium. A two-step simulation algorithm based on a gamma photon path translational invariance is developed which is subdivided into a Photon Field Building (PFB) procedure followed by a Gamma Spectrum Reconstruction (GSR) inside the detector. In section 4 the methodology is validated against gamma measurements acquired at a test field in the framework of a precision agriculture experiment. In section 5 ready-to-use formulae for inferring soil water content from proximal gamma-ray spectroscopy measurements are provided and the reliability of the method is assessed by means of an internal validation test. Finally, section 6 summarizes the main results of the work.

**2. Spatial horizons of proximal gamma-ray spectroscopy**

Proximal gamma-ray spectroscopy investigates high energy gamma radiation produced in the decays of $^{40}$K and daughter products of the $^{238}$U and $^{232}$Th decay chains, which are the only naturally occurring radionuclides producing gamma radiation of sufficient energy and intensity to be measured in the framework of in-situ surveys. Since each gamma decay has a specific emission energy, it is possible to recognize distinctive structures (photopeaks) in a gamma spectrum, which allow for the quantification of $^{40}$K, $^{238}$U and $^{232}$Th abundances in the soil source. The integrated numbers of events inside the energy ranges associated to the main photopeaks (IAEA, 2003) are typically adopted for determining the corresponding counts per second (cps) which are related to $^{40}$K, $^{238}$U and $^{232}$Th activities in the soil by some sensitivity calibration factors. While $^{208}$Tl ($^{232}$Th decay chain) and $^{40}$K are distributed solely in the soil, gamma radiation produced by the decay of $^{214}$Bi ($^{238}$U decay chain) comes both from $^{214}$Bi in the soil and from $^{214}$Bi in the atmosphere originated by the decay of $^{222}$Rn gas exhaled from rocks and soils.

The number of net counts recorded in the photopeak centered at the gamma emission energy $E$ by a detector placed at height $h$ scales with the gamma photon flux $\Phi(h)$, which can be written as follows, assuming an infinite half-space soil volume source, a homogeneous radionuclide concentration and homogeneous soil and air materials (Feng et al., 2009):

$$\Phi(h) = \frac{A_V P_\gamma}{2\mu_s(E)} \int_0^{\pi/2} \sin\theta\, e^{-\frac{\mu_a(E)h}{\cos\theta}} d\theta \qquad (1)$$

where $A_V$ is the unit volume activity in Bq/m$^3$, $P_\gamma$ is the γ-ray intensity in number of gammas per Bq, $\mu_s(E)$ and $\mu_a(E)$ are the linear attenuation coefficients in m$^{-1}$ of soil and air, respectively, and $\theta$ is the polar angle between the detector vertical symmetry axis and one radioactive unit element in the source. Linear attenuation coefficients $\mu$ define the probability $P_0$ that a gamma travels a distance $d$ in a given material without suffering any interaction and are generally expressed as the product of the mass attenuation coefficients $\mu/\rho$ (m$^2$/kg) (which depend only on the material composition and on gamma energy) times the material density $\rho$ (kg/m$^3$):

$$P_0(E) = e^{-\mu(E)d} = e^{-\left(\frac{\mu}{\rho}(E)\right)\rho d} \qquad (2)$$

Eq. (2) is what governs gamma photon survival in traversing a given material as photon attenuation is respectively positively and negatively correlated to material density and photon energy. This is the key for understanding the lateral and vertical horizons of proximal gamma-ray spectroscopy.

The vertical field of view of a gamma-ray detector placed at height $h$ can be estimated on the basis of the gamma photon flux produced within a soil thickness $t$, which can be written according to Eq. (3), where the notation is simplified for the implicit gamma energy dependence (Feng et al., 2009):

$$\Phi(h) = \frac{A_V P_\gamma}{2\mu_s} \int_0^{\pi/2} \sin\theta\, e^{-\frac{\mu_a h}{\cos\theta}} \left[1 - e^{-\frac{\mu_s t}{\cos\theta}}\right] d\theta \qquad (3)$$

The cumulative contribution to the unscattered gamma photon flux as function of soil depth has a steeper profile for decreasing gamma energy (Figure 1a) and for increasing soil density (Figure 1b). Considering a [1.2 – 1.8] g/cm$^3$ typical range of soil densities, 95% of the unscattered gamma flux at the soil surface is produced within the first [19 – 28] cm for $^{40}$K gamma photons (E = 1.46 MeV) and within the first [24 – 36] cm for $^{208}$Tl gamma photons (E = 2.61 MeV) (Table 1).

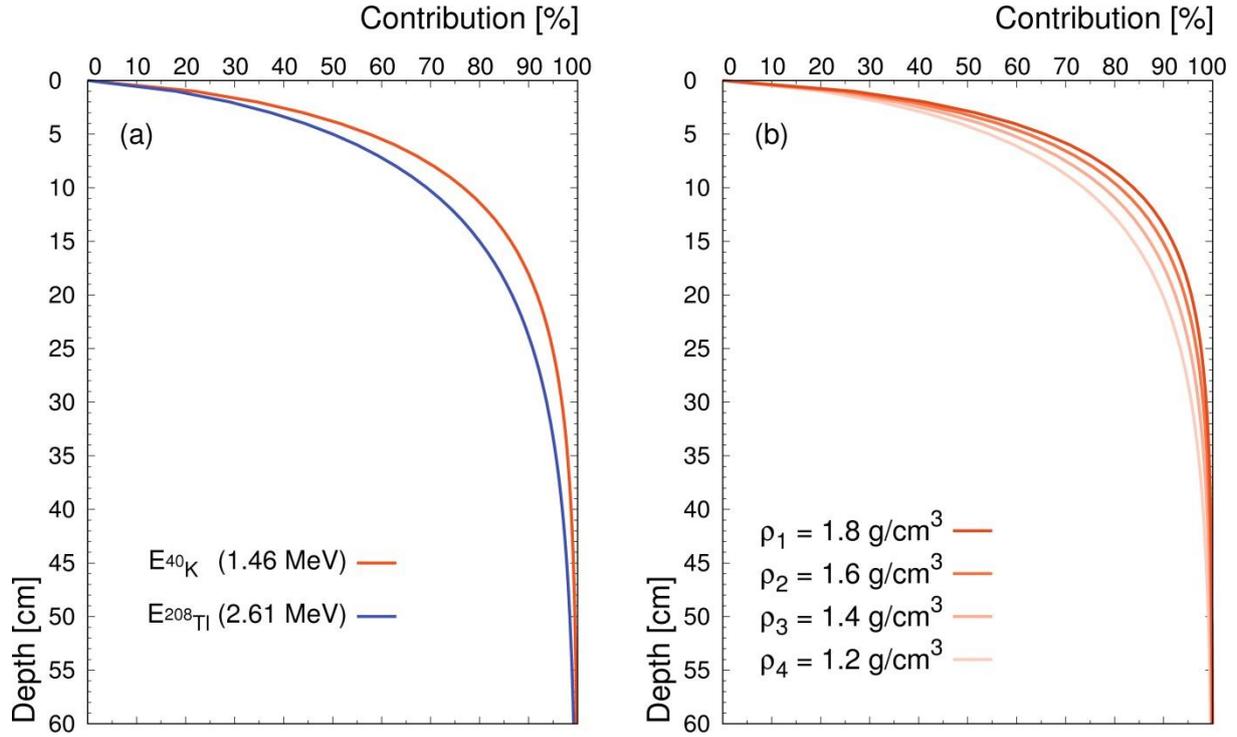

Figure 1. Cumulative percentage contribution to the unscattered gamma photon flux at ground level as function of soil depth, obtained by applying Eq. (3) with detector height $h$ equal to zero. Panel (a) refers to a soil density ($\rho$ = 1.345 g/cm$^3$) and chemical composition (Table 3) corresponding to that of the experimental site and to the $^{40}$K (1.46 MeV, $\mu/\rho$ = 0.05211 cm$^2$/g) and $^{208}$Tl (2.61 MeV, $\mu/\rho$ = 0.03874 cm$^2$/g) emission energies. Panel (b) refers to the $^{40}$K gamma emission energy and the chemical composition of the soil at the experimental site and considers a typical [1.2 – 1.8] g/cm$^3$ range of soil densities.

Table 1. Thickness of the soil layer producing 95% of the unscattered gamma photon flux at ground level for $^{40}$K (1.46 MeV) and $^{208}$Tl (2.61 MeV) gamma energies for typical values of soil bulk density.

| $\rho$ [g/cm$^3$] | Thickness [cm] | |
|---|---|---|
| | $E_{40K}$ (1.46 MeV) | $E_{208Tl}$ (2.61 MeV) |
| 1.2 | 28 | 36 |
| 1.4 | 24 | 31 |
| 1.6 | 21 | 27 |
| 1.8 | 19 | 24 |

The horizontal field of view of a gamma-ray detector placed at height $h$ can be estimated on the basis of the gamma photon flux produced within a cone of radius $r$ and opening angle $2\theta^*$ (Feng et al., 2009):

$$\Phi(h) = \frac{A_V P_\gamma}{2\mu_s} \int_0^{\theta^*} \sin\theta \, e^{-\frac{\mu_a h}{\cos\theta}} \left[ 1 - e^{-\mu_s \left( \frac{r}{\sin\theta} - \frac{h}{\cos\theta} \right)} \right] d\theta \qquad (4)$$

In the height range of proximal surveys (~ few meters), the cumulative contribution to the unscattered flux as function of the cone radius is slightly influenced by gamma energy (Figure 2a), while it sensibly changes for different heights above the ground (Figure 2b). By lifting a detector from 1 m to 10 m height the radius from which 95% of the unscattered flux is produced increases from ~15 m to ~85 m (Table 2). The differential contribution to the unscattered gamma photon flux originated by concentric hollow cylinders centered at the detector vertical axis also changes with the detector height. By increasing the detector height, the hollow cylinder providing the highest contribution is progressively farther from the detector vertical axis (Figure 3).

Gamma photon flux attenuation for increasing height is directly connected to the acquisition time which is needed for attaining a target counting statistics (Table 2): by increasing the detector height from 1 m to 10 m, the acquisition time needs to be extended by approximately 20% in order to measure the same number of events in a given energy range.

Summarizing, proximal gamma-ray spectroscopy has in principle the power of being sensitive to the physico-chemical properties of the first 30 cm of soil in an area wide fractions of hectares. The application of the method for soil water content estimation can play a strategic role in the future in filling the gap between punctual and satellite soil moisture measurements.

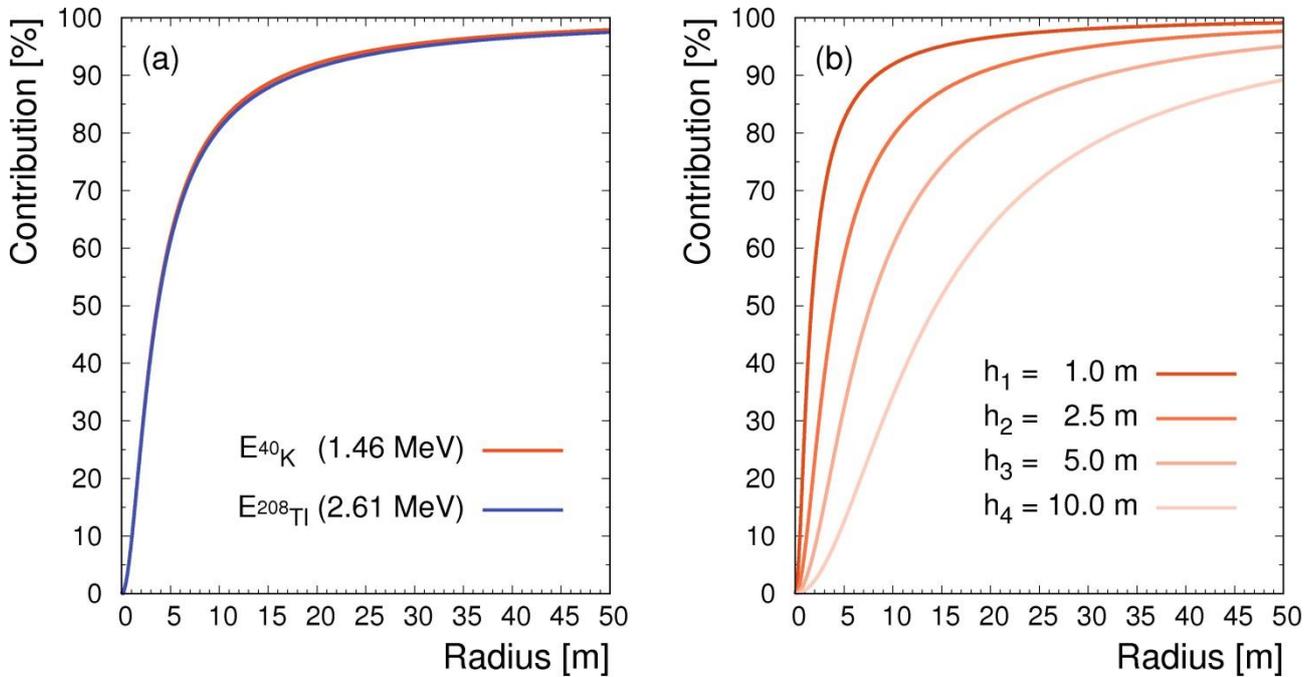

Figure 2. Cumulative percentage contribution to the unscattered gamma photon flux as function of the radial distance from the detector vertical symmetry axis obtained by applying Eq. (4) and by assuming the soil density ($\rho = 1.345$ g/cm$^3$) and chemical composition (Table 3) corresponding to that of the experimental site. Panel (a) refers to $^{40}$K (1.46 MeV) and $^{208}$Tl (2.61 MeV) emission energies and a detector height h = 2.25 m (corresponding to the height of the experimental set up). Panel b (b) refers to the $^{40}$K gamma emission energy and considers a [1 – 10] m height range.

Table 2. Radial distance from the detector vertical symmetry axis from which 95% of the unscattered gamma photon flux is produced, considering a homogeneous flat soil having density ($\rho$ = 1.345 g/cm$^3$) and chemical composition (Table 3) corresponding to that of the experimental site. Assuming unitary parent radionuclides abundances in dry soil, we report acquisition times needed to collect 10$^4$ counts for the experimental set up (see panel b) of Figure 7), estimated on the basis of the Monte Carlo method described in Section 3. Values are obtained for $^{40}$K (1.46 MeV) and $^{208}$Tl (2.61 MeV) gamma energies and considering different heights of the detector.

| Height [m] | E$_{40K}$ (1.46 MeV) | | E$_{208Tl}$ (2.61 MeV) | |
|---|---|---|---|---|
| | Radius [m] | Time [10$^3$ s] | Radius [m] | Time [10$^4$ s] |
| 1.0 | 14.9 | 1.07 | 15.7 | 2.71 |
| 2.5 | 31.4 | 1.11 | 32.9 | 2.75 |
| 5.0 | 49.6 | 1.16 | 55.5 | 2.86 |
| 10.0 | 77.5 | 1.29 | 88.5 | 3.17 |

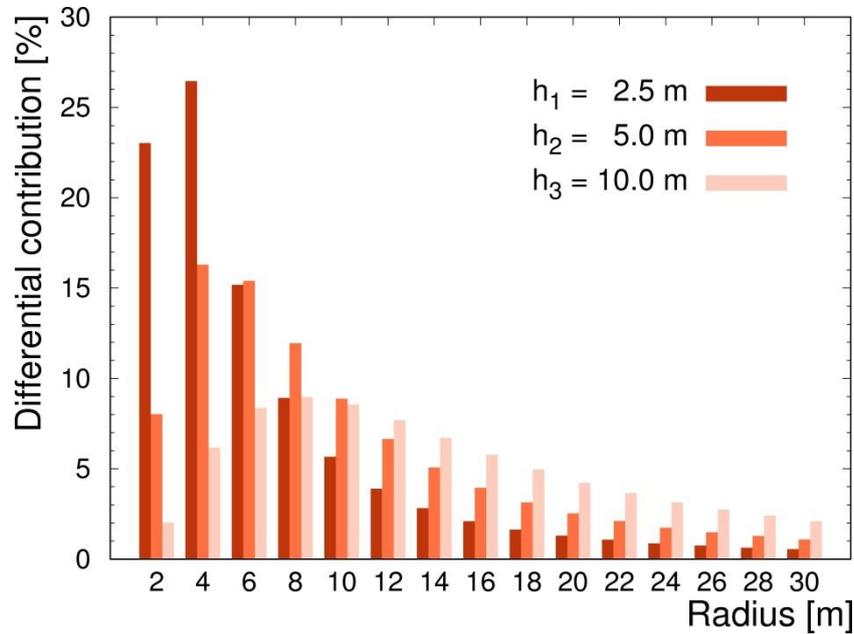

Figure 3. Histograms of the percentage contributions to the $^{40}$K (1.46 MeV) unscattered gamma photon flux produced by concentric hollow cylinders of soil centered at the detector vertical axis, where soil density ($\rho$ = 1.345 g/cm$^3$) and chemical composition (Table 3) of the experimental site were adopted. Each x axis value corresponds to the value of the external radius of each cylinder. The difference between external and internal radius of the cylinders is always 2 m. The histograms are obtained considering a 2.5 m, 5.0 m and 10.0 m height of the detector.

## 3. Gamma spectra due to a homogeneous infinite source: a Monte Carlo simulation method

The measurement of a gamma photon flux generated by a homogeneous infinite source is a well-known problem in proximal and airborne gamma-ray spectroscopy (Grasty et al., 1979). A direct Monte Carlo simulation of the gamma photon generation, propagation and detection phenomena is typically a time consuming process. This Section is focused on the description of a Monte Carlo simulation strategy thanks to which computational time issues are extremely reduced. The proposed method will be validated in a specific case-study in Section 4 and applied in Section 5.

The three major tasks of the Monte Carlo simulation applied to proximal gamma-ray spectroscopy consist in: (i) generating radioactive decays inside a source having distinct features (e.g. density, chemical composition, radionuclide distribution), (ii) chasing gamma photons by simulating their propagation/interactions in different media, (iii) reconstructing the gamma-ray spectrum recorded by a specific detection system.

A C++ Monte Carlo code based on GEANT4 (Agostinelli et al., 2003) is developed in order to perform a simulation structured into two independent steps. The adopted strategy relies on a translational invariance (Feng et al., 2009; Jacob and Paretzke, 1986; Likar et al., 2004): the first step is dedicated to the Photon Field Building (PFB) (Figure 4), while the second one is devoted to the Gamma Spectrum Reconstruction (GSR) inside the detector (Figure 5). In this context, GEANT4 is employed exclusively for gamma photons emission, propagation and tracking, while additional software was developed for the practical application of the translational invariance and of the spectral reconstruction. The translational invariance is justified by the fact that, for homogeneous traversed materials and homogeneous radioactive content of the source, a gamma radiated from a given point inside a volume element propagates equivalently to a gamma emitted at the same depth from a laterally shifted point (i.e. the two gammas undergo the same interactions and travel the same distances in all materials). As the output of the PFB process is used as input for the GSR simulation step, the simulation of gamma transport from the emission point to the detector position is completely disentangled from the simulation of the detected spectral shape. For this reason this simulation strategy is highly versatile: indeed, only the GSR process should be simulated in order to reconstruct gamma-ray spectra acquired by various detection systems for given source and traversed media.

### 3.1. Photon Field Building (PFB)

For symmetry reasons, a geometry of an infinite source and a finite-volume detection plane can be equally modeled as a geometry of a finite-volume source and an infinite detection plane (Figure 4). In the PFB process the spatial scale of the simulation is adjusted to the experimental site conditions (see section 4.1). Gammas are isotropically radiated one-by-one from homogeneously distributed emission points located inside a 1 m x 1 m x 1 m cubic source and tracked until they lose their energy down to a 0.2 MeV threshold or escape the 100 m x 100 m x 10 m global simulation volume (Figure 4a). User defined 100 m x 100 m detection planes are placed at a height 2.250 m and 3.108 m, corresponding to the height of the lower and upper surfaces of the detector container. These planes do not act as physical media in which gammas propagate and interact, but they provide the ability to record the necessary information regarding gamma states, i.e. spatial position, energy and direction cosines. The original theoretical geometry is then restored by shifting a posteriori the gamma arrival positions on the detection surface, which essentially translates into "piling up" $10^4$ $1m^2$ tiles in order to reconstruct the Photon Field Layer (PFL) that would have been obtained from a direct simulation of the real geometry (Figure 4b). In particular, symmetry reasons rule to shift gammas arrival positions in order to create a PFL having the same center and planar dimensions of the cubic source.

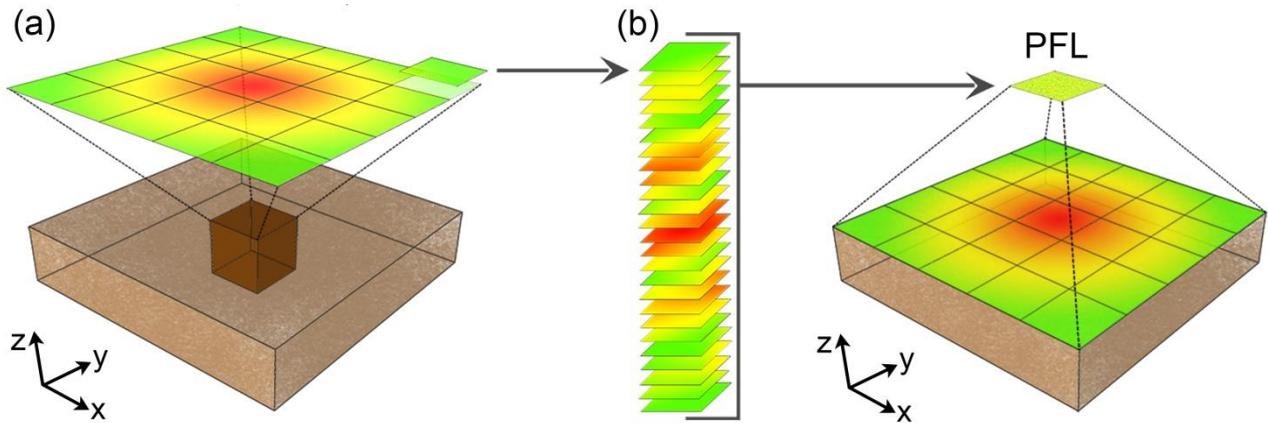

Figure 4. Scheme of the Monte Carlo Photon Field Building (PFB) process. (a) A 1 m x 1 m x 1 m cube is adopted as isotropic and homogeneous source of gamma photons which are propagated one-by-one inside the 100 m x 100 m x 10 m world volume. (b) The application of the horizontal translation to gammas arrival position on the 100 m x 100 m detection surface corresponds to a "piling up" of $10^4$ 1m$^2$ tiles of the detection plane. This procedure allows for reconstructing the Photon Field Layer (PFL), which contains information on gammas spatial position, energy and direction cosines and which corresponds to the gamma field that would have been obtained from a direct simulation of the actual source-detector configuration.

As expected, the event and energy surface densities associated to the detection surfaces are maximal at the center (i.e. on the vertical with respect to the volume source position) and gradually decrease in the radial direction, directly reflecting the fact that gammas reaching positions close to the center are the ones which, on average, traveled a shorter distance and suffered fewer interactions (Figure 5a and Figure 5c). The radial pattern disappears after the application of the translational symmetry, which leads to the reconstruction of homogeneous event and energy areal distributions on the PFL (Figure 5b and Figure 5d).

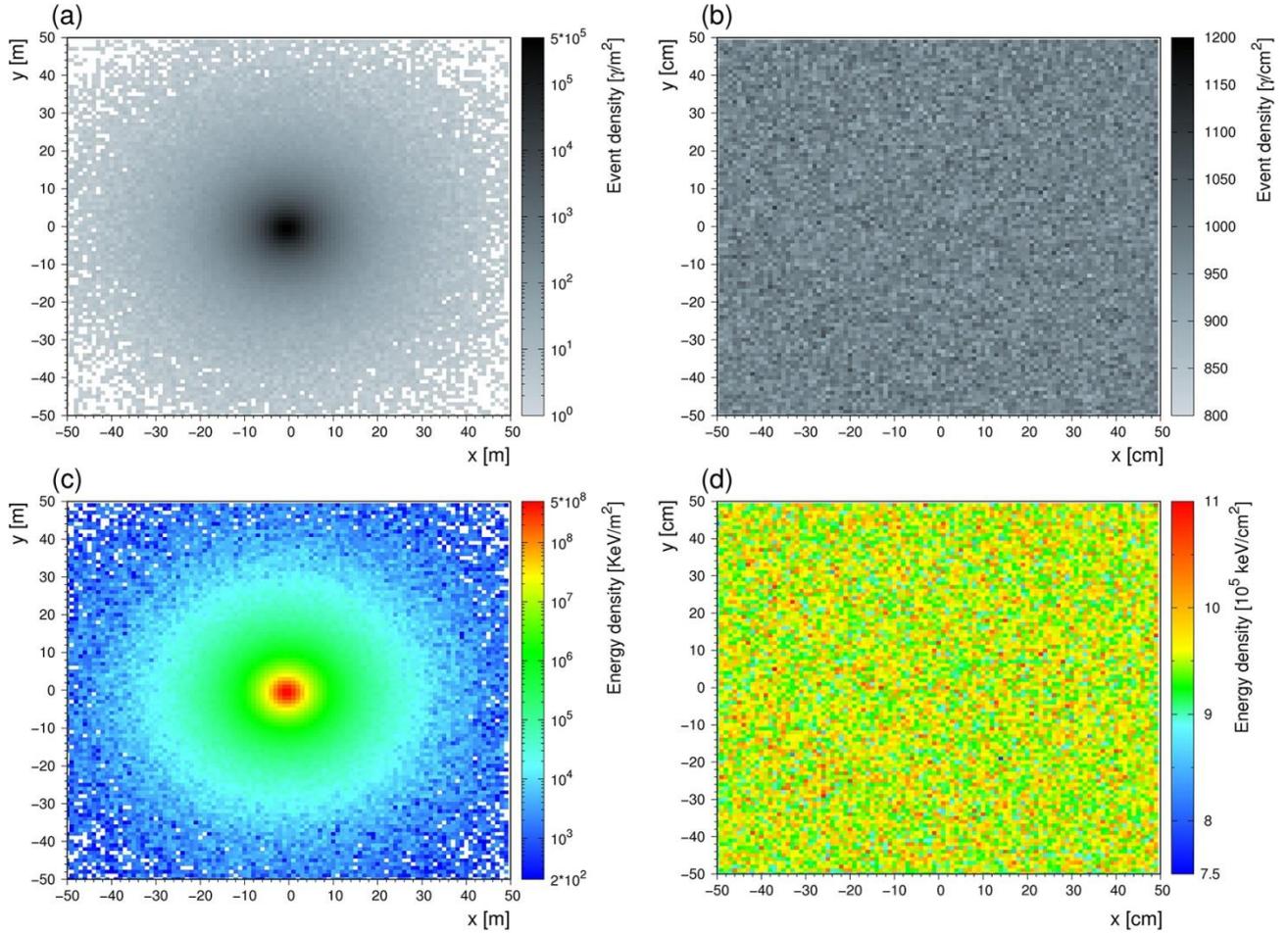

Figure 5. (a) and (c) show the gamma and energy areal distributions on the $10^4$ m$^2$ planar detection surface placed at 2.250 m height, corresponding to the height of the lower surface of the detector container placed at the experimental site. (b) and (d) illustrate the corresponding homogeneous event and energy areal distributions in the 1m$^2$ PFL obtained after the application of the translational invariance.

The most relevant variables that can be set are: the geometrical dimensions of the global system, the materials used (in terms of both chemical composition and density), the source configuration (e.g. point-like or diffuse, isotropic or collimated), the radionuclide species emitting gamma radiation and the radionuclide distribution in the source. The radioactivity of the source is defined by setting the total number of generated gammas and their energy according to the emission spectrum of the parent nuclide. Therefore, simulation of gammas emitted from natural radioactive sources is separately performed for $^{40}$K, $^{238}$U and $^{232}$Th, which is also a key point for the reconstruction of the detector fundamental spectra (Figure 8).

The number of gammas emitted per second by a unitary concentration of the i-th atomic species ($n_i$) can be determined as stated by:

$$n_i = N_i a_i \rho_{soil} V \tag{5}$$

where $N_i$ is the number of gamma photons emitted per decay by the i-th atomic species, $a_i$ is the specific activity associated to a unitary concentration of the i-th atomic species (Bq/kg) (IAEA, 2003), $\rho_{soil}$ is the soil density (kg/m$^3$) and $V$ is the source volume (m$^3$). The total number of gamma photons $\gamma_i$, emitted on average by a soil having radioelement concentration $c_i$ during a time interval $t$, can be evaluated by using:

$$\gamma_i = n_i c_i t \tag{6}$$

Storing particle direction cosines allows for distinguishing, for each detection surface, gammas traveling upwards from those traveling downwards.

### 3.2. Gamma Spectrum Reconstruction (GSR)

The PFLs of gamma photons moving upwards and downwards at respectively 2.250 m and 3.108 m height are used as inputs for the GSR stage (Figure 6a) and placed respectively on the bottom and on the top of the Monte Carlo detector prior to resuming the simulation of gamma propagation and interaction with the equipment materials. The detector employed in the experimental site and described in section 4.1 is modeled in the GSR process according to the simplified geometrical scheme shown in Figure 7. In particular, the simulated components are the detector container, the photomultiplier tube, the 1L NaI (Tl) scintillator and the detector casing. The detector behaves as a device having ideal energy resolution: what is simulated are the energy depositions inside the different detector materials, which implies that the photopeaks corresponding to specific gamma emissions are reproduced in the Monte Carlo spectrum as "Dirac delta functions" (Figure 6b). The $^{40}$K, $^{238}$U and $^{232}$Th energy-deposition spectra are broadened by folding with Gaussian resolution functions characterized by energy dependent values of the Full Width at Half Maximum (FWHM).

The shapes of six prominent photopeaks observed in measured spectra, associated with the most intense gamma lines of the $^{238}$U decay chain (351 keV from $^{214}$Pb and 609 keV, 1120keV and 1765 keV from $^{214}$Bi), of the $^{232}$Th decay chain (2614 keV from $^{208}$Tl) and to the single 1460 keV $^{40}$K gamma emission, are fitted according to a Gaussian shape, providing a mean value and a FWHM value. The FWHM values have subsequently been fitted to model the FWHM energy resolution curve (Figure 6c) according to the following simplified parameterization:

$$FWHM = k\sqrt{E} \tag{7}$$

For each radioelement, the corresponding spectrum is created as the sum of the two broadened spectra associated to gammas moving upwards and downwards (Figure 6d).

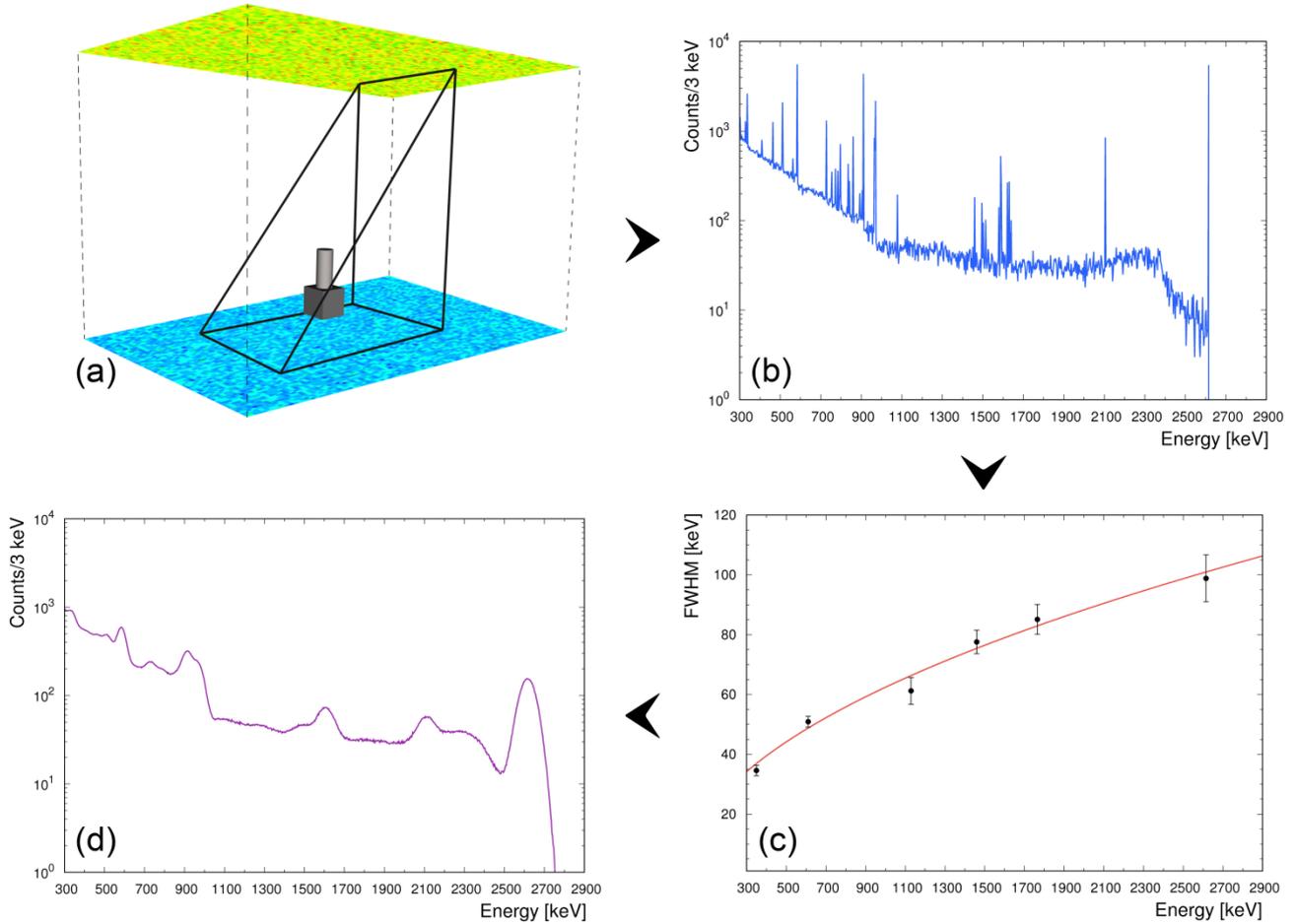

Figure 6. Scheme of the four steps of the Gamma Spectrum Reconstruction (GSR) process. (a) The detector modeled to reproduce the set up installed at the experimental field (see Figure 7b)) is placed in between the 2.250 m and 3.108 m PFLs. (b) $^{232}$Th Monte Carlo spectrum with ideal energy resolution corresponding to the gamma energy deposition obtained after the simulation of interaction of photons populating the PFLs with the modeled detector. (c) Experimental Full Width at Half Maximum (FWHM) energy resolution curve of the NaI detector installed at the experimental field obtained by fitting, according to Eq. (7) (k = 1.97 $\sqrt{keV}$), six values determined by reconstructing the Gaussian shape of prominent photopeaks observed in measured spectra. (d) $^{232}$Th spectrum obtained by summing the two Monte Carlo spectra folded with the experimental energy resolution curve and associated to the 2.250 m and 3.108 m PFLs containing respectively gammas moving upwards and downwards.

## 4. Validation of the method at an agricultural experimental field

The Monte Carlo simulation method presented above is validated in the context of a proximal gamma-ray spectroscopy experiment (Baldoncini et al., 2019; Strati et al., 2018). In section 4.1 we describe an ad-hoc NaI measurement station which was designed and installed at an agricultural experimental field with the aim of estimating soil water content on the basis of temporal changes in photopeak counting rates. Simulated K, U and Th fundamental spectra are presented in section 4.2 and adopted to validate the Monte Carlo simulation method against experimental measurements acquired at the test field.

### 4.1. Experimental site and setup

The experimental site is a 40 m x 108 m testing field (44.57° N, 11.53° E; 16 m above sea level) of the Acqua Campus, a research center of the Emiliano Romagnolo Canal (CER) irrigation district in the Emilia-Romagna region (Italy). The soil is characterized by a dry density of 1.345 g/cm$^3$ and by a loamy texture, determined on the basis of measured percentages of sand (45%), silt (40%) and clay (15%) (Strati et al., 2018). Percentages of the major oxides, quantified after a mineralogical analysis, are adopted for modeling the composition of the simulated soil material (Table 3).

Table 3. Chemical composition of the soil adopted in the Monte Carlo simulation obtained after a mineralogical analysis. Trace elements and organic matter were considered negligible for the purpose of the simulation. The $H_2O$ mass fraction refers to the structural water, corresponding to water incorporated in the formation of soil minerals.

| Major oxides/compounds | Mass Fraction [%] |
|---|---|
| $Si_2O$ | 60.41 |
| $Al_2O_3$ | 12.72 |
| $CaO$ | 10.43 |
| $Fe_2O_3$ | 4.71 |
| $MgO$ | 3.08 |
| $K_2O$ | 2.25 |
| $Na_2O$ | 1.04 |
| $TiO_2$ | 0.55 |
| $P_2O_5$ | 0.29 |
| $MnO$ | 0.13 |
| $H_2O$ | 4.34 |
| Air | 0.05 |

A total of 16 soil samples are collected within a 15 m radial distance from the detector vertical axis to homogeneously cover the area generating about 85% of the signal (Figure 4 of (Baldoncini et al., 2019)). The radioactive content of the samples is characterized on the basis of 1 hour gamma spectra acquired by the MCA_RAD system, which is made up of two coaxial HPGe detectors, able to automatically perform up to 24 measurements without human attendance (Xhixha et al., 2013). The mean abundances obtained by averaging over all samples are $a_K = (1.59 \pm 0.17)\ 10^{-2}$ g/g, $a_U = (2.48 \pm 0.25)$ µg/g and $a_{Th} = (9.37 \pm 1.12)$ µg/g. The relatively low standard deviations highlight a homogeneous radionuclide distribution over the area of interest of the experimental site.

Gamma-ray spectra are measured by a permanent gamma station specifically designed and built for the experiment (Figure 7). A 1L sodium iodide (NaI) crystal is placed inside a steel box mounted on top of a 2.25 m high steel pole. The gamma spectrometer is coupled to a photomultiplier tube base which output is processed by a digital multi-channel analyzer (MCA, CAEN γstream) having 2048 acquisition channels; the whole system is powered by a solar panel. A dedicated software is developed to post-process the output list mode files (i.e. a continuous logging of individual gamma photons arrival time and acquisition channel) in order to (i) generate gamma spectra corresponding to 15 minutes acquisition time, (ii) perform an energy calibration procedure, (iii) remove the spectral background and (iv) retrieve the net count rate in the main $^{40}K$, $^{214}Bi$ and $^{208}Tl$ photopeaks (Baldoncini et al., 2019).

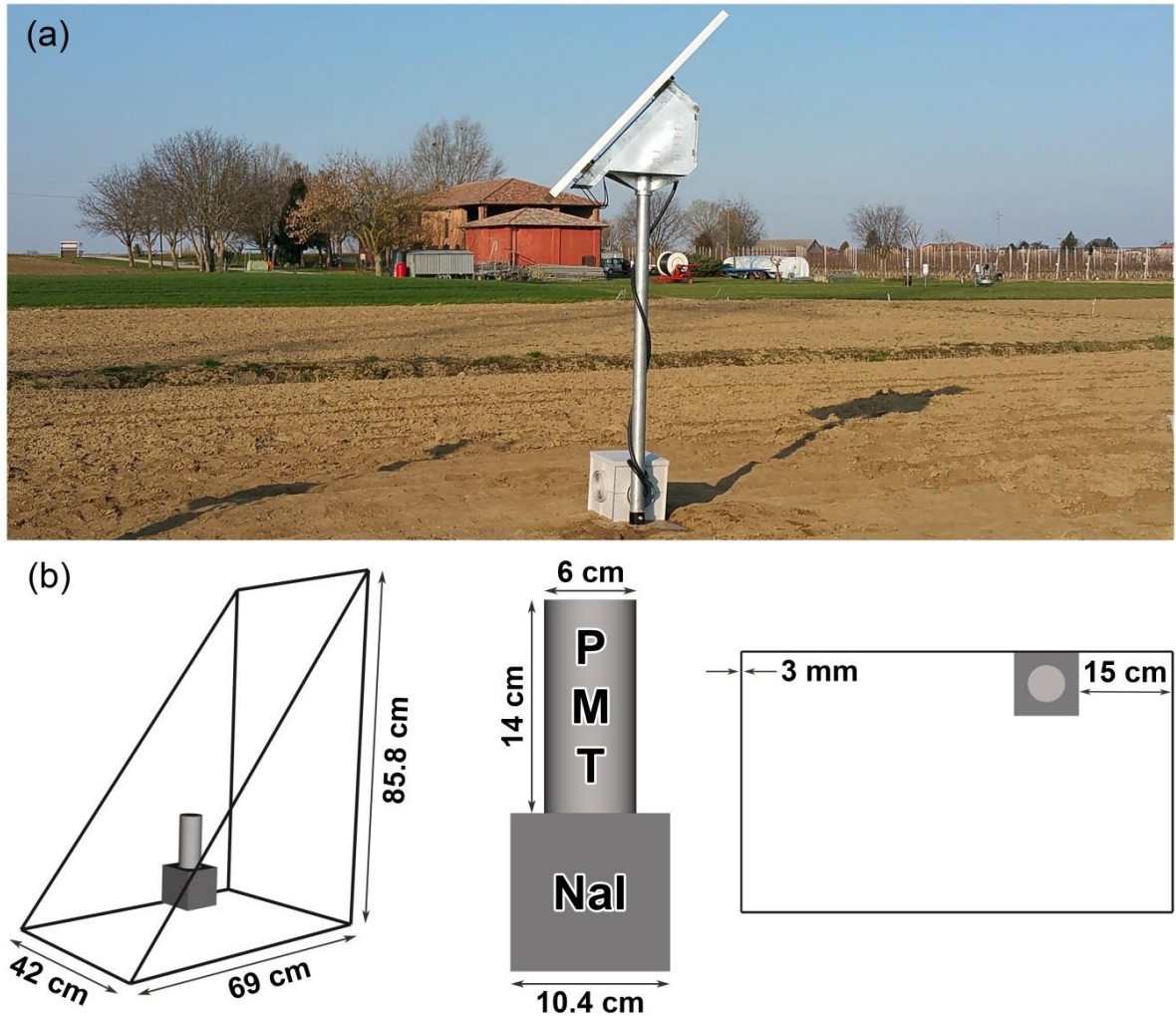

Figure 7. Panel (a) shows a picture of the NaI gamma station installed at the experimental site. Panel (b) illustrates the scheme of the experimental set up adopted for the modeling of the Monte Carlo simplified detector.

### 4.2. Simulated and experimental spectra

The Monte Carlo method illustrated in section 3 is applied to simulate gamma-ray spectra acquired at the experimental site by the setup described in section 4.1. The reliability of the simulation is tested against an experimental measurement performed in bare soil condition with known gravimetric water content (see definition in section 5.1). The weighted average gravimetric water content at calibration time $w^{Cal} = (0.163 \pm 0.008)$ kg/kg is estimated on the basis of a dedicated gravimetric measurements survey, during which soil samples are collected with the same spatial distribution adopted for the radiometric characterization (Baldoncini et al., 2019).

Independent $^{40}$K, $^{238}$U and $^{232}$Th Monte Carlo simulations are carried out in order to perform a full-spectrum detector calibration by reconstructing the so-called fundamental spectra (Figure 8), i.e. the individual radionuclide spectral shapes that a specific detection system would measure for unitary acquisition time and unitary radionuclide concentration in the soil (Hendriks et al., 2001). Fundamental spectra are generally determined by means of an experimental sensitivity calibration process according to which high statistics radiometric measurements performed on calibration homogeneous extended sources (calibration pads or natural calibration sites) successively undergo a least square analysis, necessary to unfold the separate $^{40}$K, $^{238}$U and $^{232}$Th spectral components (Caciolli et al., 2012). As the least square analysis does not intrinsically comprise any constraint on the physical gamma emission lines, this method can give rise to residual interferences in the fundamental spectral shapes of different radionuclides, especially in correspondence of photopeaks structures. In this perspective, the Monte Carlo simulation has the remarkable advantage of avoiding any type of cross-talk effect in the reconstruction of individual spectral shapes, typically caused by the minimization procedure or by the co-presence of different radionuclides having close energy gamma lines (e.g. in the case of the 0.583 MeV ($^{208}$Tl), 0.609 MeV ($^{214}$Bi) and 0.662 MeV ($^{137}$Cs) gamma emissions).

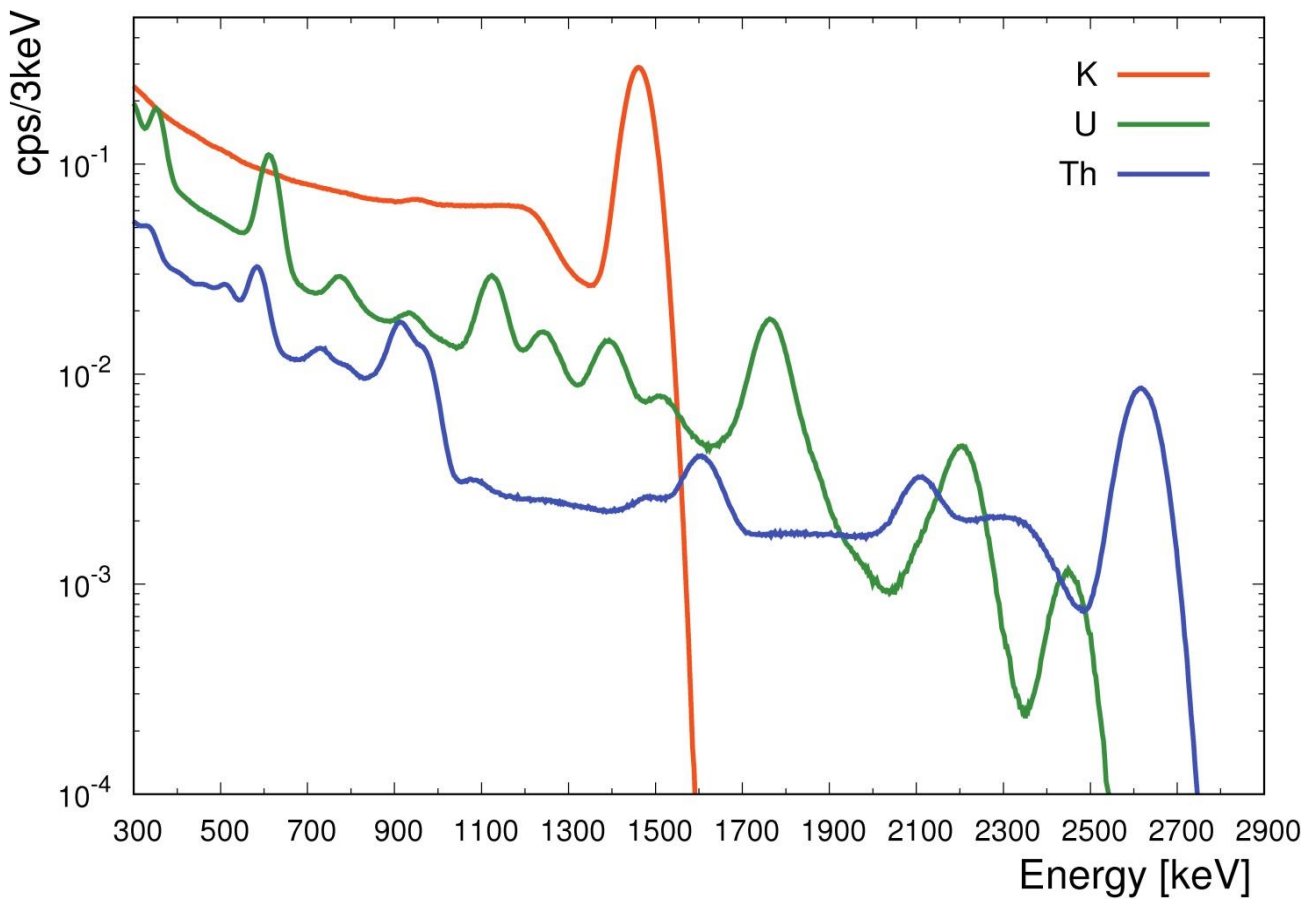

Figure 8. $^{40}$K (orange), $^{238}$U (green) and $^{232}$Th (blue) fundamental spectra obtained with the Monte Carlo simulations referred to unitary radionuclide concentrations in dry soil condition ($a_K = 10^{-2}$ g/g, $a_U = 1$ μg/g and $a_{Th} = 1$ μg/g) and to calibration gravimetric water content ($w^{Cal} = 0.163$ kg/kg).

The Full Spectrum Analysis (FSA) with Non Negative Least Squares (NNLS) fits almost the full energy spectrum of an experimental measurement by a linear combination of the fundamental spectra with the constrain of providing non negative radionuclide abundances (Caciolli et al., 2012). The fundamental spectra shown in Figure 8 are used for the FSA-NNLS analysis to reconstruct the 2 hour (10.00 a.m. – 12.00 a.m.) calibration experimental measurement concomitant with the gravimetric sampling (Figure 9). The acquisition is distant from rainfall events and scheduled irrigations and is performed with stable atmospheric parameters (i.e. air temperature and pressure, wind direction and speed).

As the adopted Monte Carlo simulation method does not structurally provide any background radiation contribution, a cosmic background spectral shape to be subtracted from the experimental measurement is inferred from a 24 hour background calibration measurement according to the approach described in (Baldoncini et al., 2017). Without the introduction of any arbitrary rescaling factor, a good agreement between experimental and simulated spectra is obtained concerning both absolute counting statistics and the spectral shape profile (Figure 9). The reconstructed $a_K = 1.63 \cdot 10^{-2}$ g/g and $a_{Th} = 10.92$ μg/g abundances are respectively compatible at 0.2 and 1.4 σ level with the radioactive content of the experimental site (see section 4.1), while the $a_U = 3.91$ μg/g abundance is highly affected by the extra contribution in the experimental measurement due to atmospheric radon.

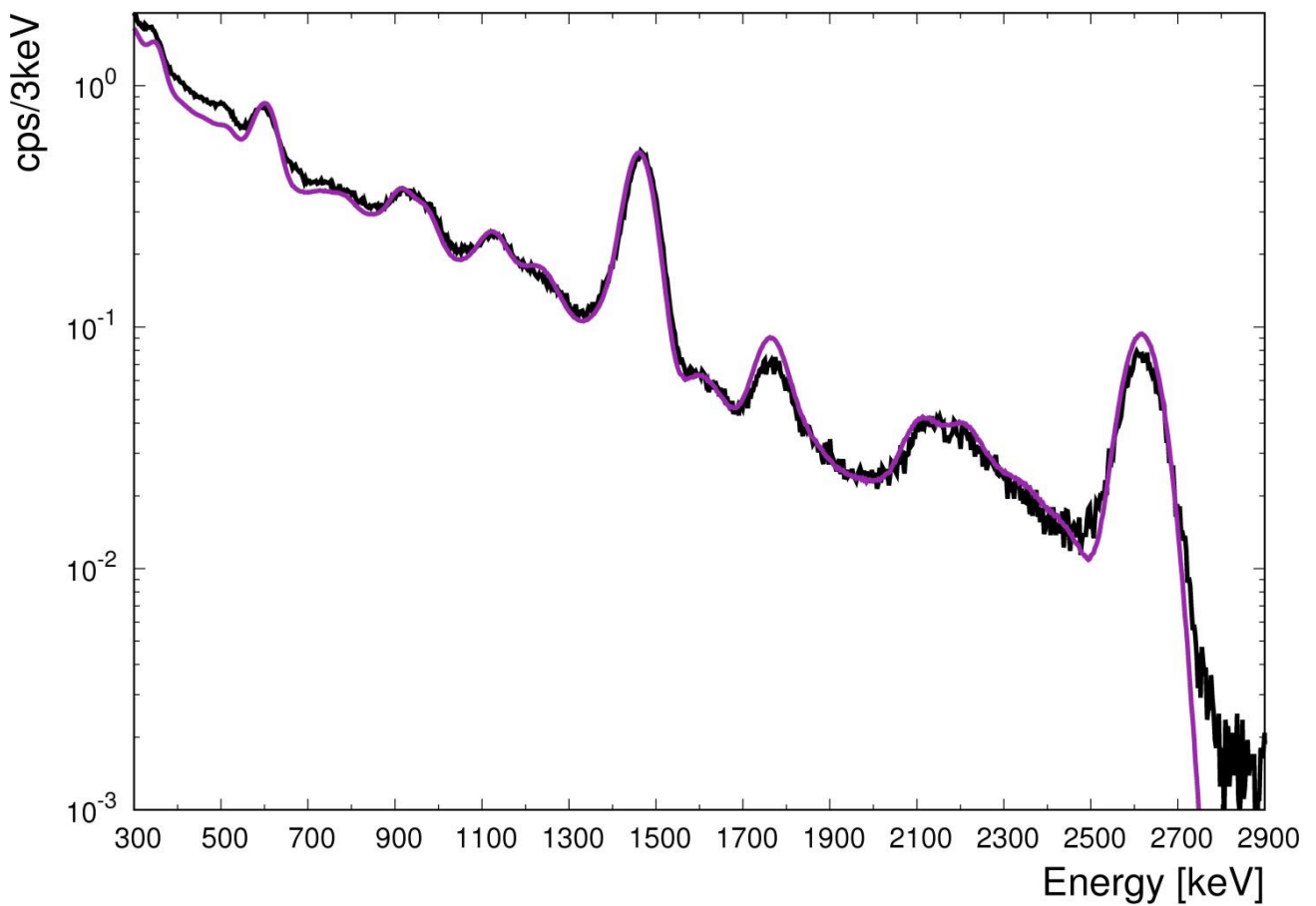

Figure 9. Gamma spectrum (black curve) acquired during the calibration day ($w^{Cal}$ = (0.163 ± 0.008) kg/kg) and reconstructed spectrum (purple curve) obtained by applying the FSA-NNLS analysis with the simulated fundamental spectra. The reconstructed abundances are $a_K = 1.63 \cdot 10^{-2}$ g/g, $a_U = 3.91$ μg/g and $a_{Th} = 10.92$ μg/g.

## 5. Application of the method to soil water content estimation in precision agriculture

As the objective of this study is investigating the potentialities of proximal gamma-ray spectroscopy in assessing soil water content for precision agriculture, it is of fundamental relevance inquiring into the possibility of distinguishing water distributed in the soil matrix pores from that incorporated in the formation of minerals. Indeed, addressing this question is of decisive importance since the plant available water, i.e. the store of soil water readily available to plants for transpiration and consequently growth, is typically just a fraction of the water mass contained in soil pores. In Section 3 a Monte Carlo method based on advanced simulation and software tools was presented: however, this know-how generally does not belong to the expertise of the community involved in the field of precision agriculture. Therefore, Section 5.1 and Section 5.2 are devoted to the development of a ready-to-use general recipe to be employed for inferring soil water content from gamma-ray spectroscopy measurements without the need of any custom Monte Carlo simulation. Finally, in Section 5.3 an internal validation test is performed with the aim of assessing the reliability of the method.

## 5.1. A recipe for inferring soil water content

Soil is a complex system made up of a heterogeneous mixture of solid, liquid and gaseous phases. In the presence of a mixture, the mass attenuation coefficient $(\mu/\rho)$ referred to a given gamma emission energy (Section 2) is obtained as the mass abundance weighted sum of the mass attenuation coefficients of individual soil material constituents. Gamma-ray spectroscopy essentially treats soil as a two-phase medium in which the total mass $M$ comprises a mass of solid constituents $M^S$ (the largest portions typically due to the oxides $SiO_2$, $Al_2O_3$ and $CaO$) and a water mass $M_{H2O}$. Considering that in the energy range of natural gamma emissions typically monitored in radiometric measurements (~1 MeV) mineral elements have comparable mass attenuation coefficients (Lovborg, 1984) which are significantly different from that of water, in the expression for the mass attenuation coefficient it is possible to split a water component and a solid component as follows:

$$\left(\frac{\mu}{\rho}\right) = \frac{M^S}{M}\left(\frac{\mu}{\rho}\right)_S + \frac{M_{H2O}}{M}\left(\frac{\mu}{\rho}\right)_{H2O} \tag{8}$$

where $(\mu/\rho)_s$ and $(\mu/\rho)_{H2O}$ correspond respectively to the mass attenuation coefficient of the soil solid portion and of water.

In the mentioned energy range, the dominating gamma-ray interaction is Compton scattering, which cross section is essentially proportional to the electron density $Z/A$. As soil major constituents ($Z<30$) have $Z/A$ values close to 0.5 and considering that water has a fixed $Z/A = 0.556$, it turns out that the typical value for the ratio between the mass attenuation coefficient of the solid portion to that of water is 0.90, corresponding to the value adopted by (Carroll, 1981; Grasty, 1997) in the processing of $^{40}$K and $^{208}$Tl photopeak signals from airborne radiometric surveys.

Generally speaking, soil water content $w$ at time $t$ can be inferred from gamma-ray spectroscopy measurements, provided a detector calibration on the basis of a soil water content independent calibration measurement. The key for soil water content assessment is the ratio between the gamma signal measured at the calibration time $S^{Cal}$ [cps], for which the soil water content $w^{Cal}$ [kg/kg] is known, and the gamma signal measured at the time $t$ $S(t)$ [cps]. By considering that the gamma signal measured in a given photopeak energy is directly proportional to the parent radionuclide abundance and inversely proportional to the soil mass attenuation coefficient (as can be inferred from Eq. (8)), the following equation for soil water content can be derived:

$$w(t) = \frac{S^{Cal}}{S(t)} \cdot \left(0.90 + w^{Cal}\right) - 0.90 \tag{9}$$

where soil water content $w$ [kg/kg] is here defined as the water-to-dry fraction, i.e. the ratio between the soil water mass and the solid constituents mass:

$$w = \frac{M_{H2O}}{M_S} \tag{10}$$

Soil water mass can be distinguished into mass of water filling soil matrix pores ($M_{H2O}^P$) and mass of structural water ($M_{H2O}^{struct}$), i.e. water incorporated in the formation of soil minerals. Distinguishing soil water content on the basis of water allocation allows for focusing onto two distinct aspects. From one side it is relevant to recognize that precision farming aimed at water resources optimization deals with water distributed in soil pores. Secondly, it becomes clear that, according to the definition of soil water content given in Eq. (10), Eq. (9) cannot be calibrated by means of traditional soil gravimetric measurements. Indeed, as the latter are typically performed by drying the sample at ~105 °C for 24 hours (Hillel, 1998), the dry sample will still comprise the structural water mass which is not lost as such low heating temperatures. In this perspective, we developed ready-to-use general formulae that can be easily adopted in precision agriculture and which can be, at the same time, easily calibrated:

$$w_{G^{40}K}(t) = \frac{S^{Cal}_{^{40}K}}{S_{^{40}K}(t)} \cdot \left[(0.903 \pm 0.011) + w_G^{Cal}\right] - (0.903 \pm 0.011) \tag{11}$$

$$w_{G^{208}Tl}(t) = \frac{S^{Cal}_{^{208}Tl}}{S_{^{208}Tl}(t)} \cdot \left[(0.915 \pm 0.009) + w_G^{Cal}\right] - (0.915 \pm 0.009) \tag{12}$$

Eq. (11) and Eq. (12) respectively provide the gravimetric soil water content inferred from $^{40}K$ and $^{208}Tl$ photopeak gamma signals: the $S^{Cal}$ [cps] and $S$ [cps] terms are, for each energy window, the photopeak signals recorded respectively at calibration time and at time $t$, $w_G^{Cal}$ and $w_G(t)$ are respectively the gravimetric soil water content at calibration time and at time $t$, with gravimetric soil water content defined as:

$$w_G = \frac{M_{H2O}^P}{M^S + M_{H2O}^{struct}} \tag{13}$$

## 5.2. Do structural water and chemical composition affect soil water content estimation?

The gravimetric soil water content recipes presented in Eq. (11) and Eq. (12) are derived on the basis of average structural water fractions and soil chemical compositions. In case a detailed soil mineralogical analysis is available, e.g. obtained by means of XRF measurements, a site specific formula for gravimetric soil water content can be adopted for each gamma emission energy:

$$w_G(t) = \frac{S^{Cal}}{S(t)} \cdot \left[\Omega + w_G^{Cal}\right] - \Omega = \frac{S^{Cal}}{S(t)} \cdot \left[\Psi + (1-\Psi)f_{H2O}^{struct} + w_G^{Cal}\right] - \left(\Psi + (1-\Psi)f_{H2O}^{struct}\right) \quad (14)$$

where the energy dependent adimensional factor $\Omega$ is explicitly written in terms of the energy dependent adimensional $\Psi$ parameter and of the mass fraction of structural water $f_{H2O}^{struct}$, defined as:

$$\Psi = \frac{\left(\frac{\mu}{\rho}\right)_S}{\left(\frac{\mu}{\rho}\right)_{H2O}} \quad (15)$$

and

$$f_{H2O}^{struct} = \frac{M_{H2O}^{struct}}{M^S + M_{H2O}^{struct}} \quad (16)$$

The average fraction of structural water $f_{H2O}^{struct} = \left(0.012_{-0.010}^{+0.030}\right)$ [kg/kg] adopted in Eq. (11) and Eq. (12) was obtained by considering the average (0.41 ± 0.29) fraction of structural water mass to Loss Of Ignition (LOI) mass reported in (Sun et al., 2009), combined with the median (0.03 kg/kg), 1st quartile (0.02 kg/kg) and 3rd quartile (0.058 kg/kg) LOI values reported in (Weynants et al., 2013).

In order to evaluate the two distinct average $\Omega$ factors entering in Eq. (11) and Eq. (12), mean $\Psi$ values were determined separately for the $^{40}$K and $^{208}$Tl gamma emission energy, corresponding respectively to $\Psi_{40K}$ = (0.902 ± 0.010) and $\Psi_{208Tl}$ = (0.914 ± 0.009). Mean $\Psi$ values were derived by averaging individual values referred to standard soil compositions (Table 4), in turn calculated as the ratio between the solid and water mass attenuation coefficients (see Eq. (15)). For each gamma emission energy, mass attenuation coefficient for the soil solid portion and for water were computed by applying the Beer-Lambert attenuation law to gamma photon counting results obtained from monochromatic mono-directional Monte Carlo simulations, which were performed by adopting a material of known thickness made up by 100% soil solid phase and by 100% water, respectively. The results presented in Table 4 for the soil major oxides and for standard soil compositions show that the $\Psi$ coefficient has some $Z$ dependence, i.e. a site dependence related to the specific soil chemical composition, as well as a gamma energy dependence.

Table 4: Ratio Ψ between the mass attenuation coefficient of the soil solid portion to that of water (see Eq. (15)) for the major oxides constituting the soil material and for different standard soils. Ψ values are separately given for the $^{40}$K (1.46 MeV) and $^{208}$Tl (2.61 MeV) gamma energies.

| Major oxides | Ψ($^{40}$K) | Ψ($^{208}$Tl) |
| --- | --- | --- |
| SiO$_2$ | 0.900 | 0.910 |
| Al$_2$O$_3$ | 0.884 | 0.893 |
| CaO | 0.903 | 0.927 |
| Fe$_2$O$_3$ | 0.864 | 0.894 |
| MgO | 0.895 | 0.904 |
| K$_2$O | 0.883 | 0.908 |
| Na$_2$O | 0.872 | 0.880 |
| TiO$_2$ | 0.861 | 0.882 |
| MnO | 0.844 | 0.875 |

| Standard soils | Ψ($^{40}$K) | Ψ($^{208}$Tl) |
| --- | --- | --- |
| Experimental site[1] | 0.895 | 0.908 |
| Beck[2] | 0.906 | 0.916 |
| Soil 1[3] | 0.913 | 0.924 |
| Soil 5[3] | 0.889 | 0.905 |
| Soil 2[3] | 0.914 | 0.924 |
| Nist SRM 2711[4] | 0.895 | 0.907 |

[1] Table 3Table 3
[2] (Beck et al., 1972)
[3] (Jacob et al., 1994)
[4] (Mackey et al., 2010)

## 5.3. Performance of the recipe

This section is devoted to present the results of an interval validation test aimed at assessing the performances of the method for estimating soil water content from radiometric signals acquired by a proximal gamma-ray station. According to the procedure described in Section 5.2, the equations for determining the water-to-dry fractions (Eq. (10)) referred to the specific composition of the soil at the experimental site (Table 3) are derived for the $^{40}$K and $^{208}$Tl gamma-ray emission energies (see also Table 4):

$$w_{^{40}K} = \frac{S_{^{40}K}^{Cal}}{S_{^{40}K}} \cdot \left(0.895 + w^{Cal}\right) - 0.895 \tag{17}$$

$$w_{208_{Tl}} = \frac{S^{Cal}_{208_{Tl}}}{S_{208_{Tl}}} \cdot (0.908 + w^{Cal}) - 0.908 \tag{18}$$

Monte Carlo simulations of $^{40}$K and $^{208}$Tl gamma signals are carried out by adopting soil composition, dry bulk density and modeled experimental set up described in Section 4 and by varying water-to-dry fractions $w$ ranging from dry soil condition up to saturation (Table 5). For each configuration, $10^{10}$ initial number of events is simulated and the wet bulk density is increased accordingly. Linear regressions between estimated and input water-to-dry fraction $w$ show that the described method allows for determining $w$ with an uncertainty <1%. The best fit linear regression lines for water-to-dry fractions estimated from $^{40}$K and $^{208}$Tl simulated gamma signals are respectively $w^{output}_{40_K} = (0.997 \pm 0.001) \cdot w^{input} - (0.0003 \pm 0.0003)$ and $w^{output}_{208_{Tl}} = (0.993 \pm 0.001) \cdot w^{input} - (0.0006 \pm 0.0004)$, both characterized by a coefficient of determination equal to 1.

Table 5. Results of the internal validation test for the assessment of the water-to-solid mass fraction w on the basis of $^{40}$K and $^{208}$Tl simulated gamma signals. The first two columns report respectively the input w and density values ρ of the simulated soil material. The output values in terms of count rate for unitary radioelement abundance referred to dry soil condition and water-to-dry fractions w are reported for $^{40}$K and $^{208}$Tl gamma energies (see Eq. (17) and Eq. (18)).

| INPUT | | OUTPUT $^{40}$K | | OUTPUT $^{208}$Tl | |
|---|---|---|---|---|---|
| w [kg/kg] | ρ [g/cm³] | Count rate [cps] | w [kg/kg] | Count rate [cps] | w [kg/kg] |
| 0.045 | 1.345 | 9.10 | 0.045 | 0.364 | 0.045 |
| 0.094 | 1.390 | 8.66 | 0.093 | 0.346 | 0.092 |
| 0.167 | 1.449 | 8.07 | 0.166 | 0.323 | 0.165 |
| 0.261 | 1.516 | 7.41 | 0.260 | 0.297 | 0.258 |
| 0.372 | 1.583 | 6.76 | 0.371 | 0.271 | 0.369 |
| 0.571 | 1.680 | 5.85 | 0.569 | 0.235 | 0.567 |

The simulated detection system is expected to record for soil dry condition and radionuclide unitary abundances about 9.1 cps in the $^{40}$K photopeak to be compared with 0.36 cps in the $^{208}$Tl photopeak (Table 5). Since in natural contexts $^{208}$Tl gamma emission is characterized by a lower gamma luminosity compared to that of $^{40}$K and since detection efficiency decreases for increasing photon energy, comparable counting statistics in the $^{40}$K and $^{208}$Tl photopeaks for unitary radionuclide abundances are obtained by integrating over acquisition times having a ratio of about 1:25 (see also Table 2).

Starting from soil in dry condition characterized by unitary radionuclide abundances, the addition of water leads to a dilution of the radionuclide concentration, an effective increase in soil density and an almost linear scaling of the spectral shapes. By increasing the water-to-dry fraction by a factor of 5, an average 32% reduction of the bin-by-bin counting statistics for both $^{40}$K and $^{232}$Th is observed (Figure 10).

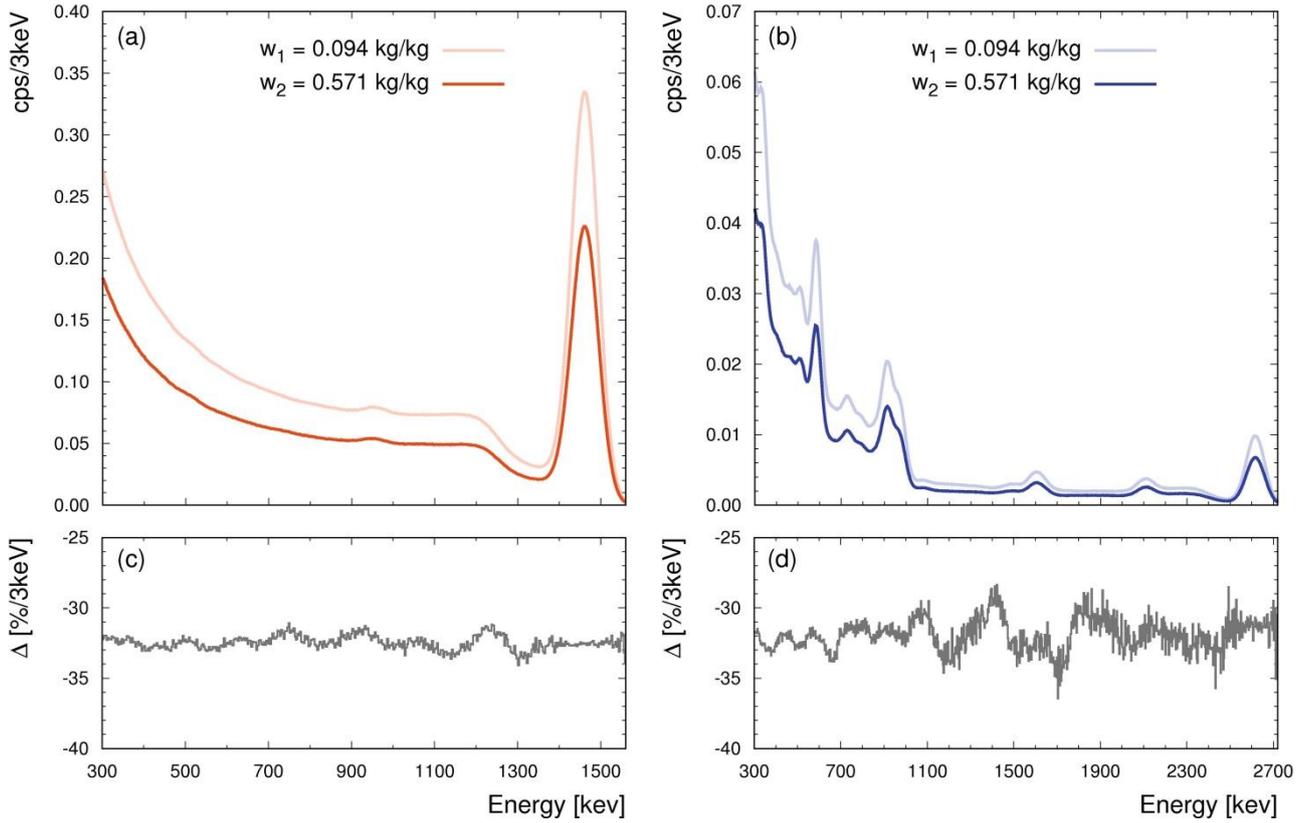

Figure 10. $^{40}$K (a) and $^{232}$Th (b) fundamental spectra obtained considering unitary radionuclide concentrations in dry soil condition and fro two distinct water-to-dry factions $w_1 = 0.094$ kg/kg and $w_2 = 0.571$ kg/kg (see Table 5). Bin-to-bin percentage differences ($\Delta$) between fundamental spectra simulated with $w_1$ and $w_2$ are shown for $^{40}$K (c) and $^{232}$Th (d). Both fundamental spectra and percentage differences are displayed adopting a 3 keV bin width.

## 6. Conclusions

Proximal gamma-ray spectroscopy is being recognized as one of the best space-time trade off methods for a continuous and non-invasive determination of soil moisture dynamics and as an extraordinary joining link between punctual and satellite fields of view. However, the potentialities of the method have not been fully explored.

In this paper a Monte Carlo method is applied to the simulation of NaI gamma-ray spectra for soil water content estimation at field scale. The strength of this approach relies in the adoption of a two-steps strategy obtained by splitting the simulation into an equipment-independent Photon Field Building (PFB), which simulates gamma transport from the source to the detector position, and a Gamma Spectrum Reconstruction (GSR) process, devoted to the simulation of the employed detection system and of the recorded gamma spectra. This method allows for (i) the achievement of high simulated counting statistics with the potential for real time processing, (ii) calibrate for fundamental spectra produced by individual radionuclides, (iii) perform sensitivity studies for distinct environmental variables, e.g. soil moisture. In the perspective of investigating variegated experimental scenarios, the high degree of customization provides an effective tool for feasibility and sensitivity studies. Different environmental conditions related to physical and chemical variables, distinct detection set ups and fields of view can be simulated.

The reliability of the method is effectively validated with gamma spectra measured by a permanent station installed at an agricultural experimental site, which is constituted by a 1L NaI detector placed at a height of 2.25 m, sensitive to an area having a ~25 m radius and to a depth of approximately 30 cm.

The developed theoretical model which relates soil water content to gamma signal according to an inverse proportionality law needs, in addition to the signal and water content values at calibration time, a soil dependent coefficient $\Omega$. The energy dependent adimensional coefficient $\Omega$ combines the amount of structural water and the ratio $\Psi$ between the mass attenuation coefficient of the soil solid portion to that of water. The latter, determined by a Monte Carlo approach, is provided for the major soil oxides, which can be combined according to their mass abundance for calculating site specific $\Psi$ values. The $\Omega$ coefficient is provided both for the specific composition of the experimental site and for standard soils.

The theoretical model is applied in the framework of a Monte Carlo synthetic calibration, providing an excellent agreement in terms of linear regression between input and output soil water contents, inferred from simulated $^{40}$K and $^{208}$Tl gamma signals. By simulating $10^{10}$ initial events in the soil source, the ~$10^6$ reconstructed statistics inside the detector is affected by an uncertainty < 0.1%. The excellent results in terms of slope, intercept and coefficient of determination values demonstrate the capability of the proposed method in estimating soil water content with an average uncertainty < 1%.

**Acknowledgements**

This work was partially founded by the National Institute of Nuclear Physics (INFN) through the ITALian RADioactivity project (ITALRAD) and by the Theoretical Astroparticle Physics (TAsP) research network. The authors would like to acknowledge the support of the Project Agroalimentare Idrointelligente CUP D92I16000030009, of the Geological and Seismic Survey of the Umbria Region (UMBRIARAD), of the University of Ferrara (Fondo di Ateneo per la Ricerca scientifica FAR 2016) and of the MIUR (Ministero dell'Istruzione, dell'Università e della Ricerca) under MIUR-PRIN-2012 project.

The authors thank the staff of GeoExplorer Impresa Sociale s.r.l. for their support and Renzo Valloni, Vincenzo Guidi, Barbara Fabbri, Stefano Anconelli, Domenico Solimando, Enrico Calore, Sebastiano Fabio Schifano, Raffaele Tripiccione, Claudio Pagotto and Ivan Callegari for their collaboration which made possible the realization of this study. The authors show their gratitude to Marco Bittelli, Giovanni Fiorentini, Miguel García Castaño, Ilaria Marzola, Michele Montuschi, Barbara Ricci and Carlos Rossi Alvarez for useful comments and discussions.

The authors thank the Università degli Studi di Ferrara and INFN-Ferrara for the access to the COKA GPU cluster.

**References**

Agostinelli, S., Allison, J., Amako, K.a., Apostolakis, J., Araujo, H., Arce, P., Asai, M., Axen, D., Banerjee, S., Barrand, G., 2003. GEANT4—a simulation toolkit. Nuclear instruments and methods in physics research section A: Accelerators, Spectrometers, Detectors and Associated Equipment 506, 250-303.

Allyson, J., Sanderson, D., 1998. Monte Carlo simulation of environmental airborne gamma-spectrometry. Journal of Environmental Radioactivity 38, 259-282.

Androulakaki, E., Kokkoris, M., Tsabaris, C., Eleftheriou, G., Patiris, D., Pappa, F., Vlastou, R., 2016. In situ γ-ray spectrometry in the marine environment using full spectrum analysis for natural radionuclides. Applied Radiation and Isotopes 114, 76-86.

Baldoncini, M., Albèri, M., Bottardi, C., Chiarelli, E., Raptis, K.G.C., Strati, V., Mantovani, F., 2019. Biomass water content effect in soil water content assessment via proximal gamma-ray spectroscopy. Geoderma, 335, 69-77.

Baldoncini, M., Albéri, M., Bottardi, C., Minty, B., Raptis, K.G., Strati, V., Mantovani, F., 2017. Airborne gamma-ray spectroscopy for modeling cosmic radiation and effective dose in the lower atmosphere. IEEE Transactions on Geoscience and Remote Sensing.

Beamish, D., 2013. Gamma ray attenuation in the soils of Northern Ireland, with special reference to peat. Journal of environmental radioactivity 115, 13-27.

Beamish, D., 2014. Peat mapping associations of airborne radiometric survey data. Remote Sensing 6, 521-539.

Beamish, D., 2015. Relationships between gamma-ray attenuation and soils in SW England. Geoderma 259, 174-186.

Beck, H.L., Gogolak, C., DeCampo, J., 1972. In situ Ge (Li) and NaI (Tl) gamma-ray spectrometry. CM-P00066834.

Bogena, H.R., Huisman, J.A., Güntner, A., Hübner, C., Kusche, J., Jonard, F., Vey, S., Vereecken, H., 2015. Emerging methods for noninvasive sensing of soil moisture dynamics from field to catchment scale: A review. Wiley Interdisciplinary Reviews: Water 2, 635-647.

Bristow, Q., 1983. Airborne γ-ray spectrometry in uranium exploration. Principles and current practice, Nuclear Geophysics. Elsevier, pp. 199-229.

Brocca, L., Crow, W.T., Ciabatta, L., Massari, C., de Rosnay, P., Enenkel, M., Hahn, S., Amarnath, G., Camici, S., Tarpanelli, A., 2017. A review of the applications of ASCAT soil moisture products. IEEE Journal of Selected Topics in Applied Earth Observations and Remote Sensing 10, 2285-2306.

Caciolli, A., Baldoncini, M., Bezzon, G., Broggini, C., Buso, G., Callegari, I., Colonna, T., Fiorentini, G., Guastaldi, E., Mantovani, F., 2012. A new FSA approach for in situ γ ray spectroscopy. Science of the Total Environment 414, 639-645.

Carroll, S.S., Carroll, T.R., 1989. Effect of forest biomass on airborne snow water equivalent estimates obtained by measuring terrestrial gamma radiation. Remote Sensing of Environment 27, 313-319.

Carroll, T., 1981. Airborne Soil Moisture Measurement Using Natural Terrestrial Gamma Radiation. Soil Science 132, 358-366.

Chirosca, A., Suvaila, R., Sima, O., 2013. Monte Carlo simulation by GEANT 4 and GESPECOR of in situ gamma-ray spectrometry measurements. Applied Radiation and Isotopes 81, 87-91.

Coulouma, G., Caner, L., Loonstra, E.H., Lagacherie, P., 2016. Analysing the proximal gamma radiometry in contrasting Mediterranean landscapes: Towards a regional prediction of clay content. Geoderma 266, 127-135.

De Groot, A., Van der Graaf, E., De Meijer, R., Maučec, M., 2009. Sensitivity of in-situ γ-ray spectra to soil density and water content. Nuclear Instruments and Methods in Physics Research Section A: Accelerators, Spectrometers, Detectors and Associated Equipment 600, 519-523.

Dierke, C., Werban, U., 2013. Relationships between gamma-ray data and soil properties at an agricultural test site. Geoderma 199, 90-98.

Feng, T., Cheng, J., Jia, M., Wu, R., Feng, Y., Su, C., Chen, W., 2009. Relationship between soil bulk density and PVR of in situ γ spectra. Nuclear Instruments and Methods in Physics Research Section A: Accelerators, Spectrometers, Detectors and Associated Equipment 608, 92-98.

Grasty, R., Kosanke, K., Foote, R., 1979. Fields of view of airborne gamma-ray detectors. Geophysics 44, 1447-1457.

Grasty, R.L., 1997. Radon emanation and soil moisture effects on airborne gamma-ray measurements. Geophysics 62, 1379-1385.

Heggemann, T., Welp, G., Amelung, W., Angst, G., Franz, S.O., Koszinski, S., Schmidt, K., Pätzold, S., 2017. Proximal gamma-ray spectrometry for site-independent in situ prediction of soil texture on ten heterogeneous fields in Germany using support vector machines. Soil and Tillage Research 168, 99-109.

Hendriks, P., Limburg, J., De Meijer, R., 2001. Full-spectrum analysis of natural γ-ray spectra. Journal of Environmental Radioactivity 53, 365-380.

Hillel, D., 1998. Environmental soil physics: Fundamentals, applications, and environmental considerations. Elsevier, p.13.

IAEA, 2003. Guidelines for Radioelement MappingUsing Gamma Ray Spectrometry Data. International Atomic Energy Agency, Vienna. Technical ReportSeries No. 323

Jacob, P., Debertin, K., Miller, K., Roed, J., Saito, K., Sanderson, D., 1994. Report 53. Journal of the ICRU, NP-NP.

Jacob, P., Paretzke, H.G., 1986. Gamma-ray exposure from contaminated soil. Nuclear Science and Engineering 93, 248-261.

Killeen, P., 1963. Killeen, PG, Gamma ray spectrometric methods in uranium exploration–application and interpretation; in Geophysics and Geochemistry in the Search for Metallic Ores; Peter J. Hood, editor; Geological Survey of Canada, Economic Geology Report 31, p. 163-229, 1979. Economic Geology Report, 163.

Likar, A., Vidmar, T., Lipoglavšek, M., Omahen, G., 2004. Monte Carlo calculation of entire in situ gamma-ray spectra. Journal of environmental radioactivity 72, 163-168.

Loijens, H.S., 1980. Determination of soil water content from terrestrial gamma radiation measurements. Water Resources Research 16, 565-573.

Lovborg, L., 1984. Calibration of Portable and Airborne Gamma-Ray Spectrometers: Theory of Problems and Facilities.

Mackey, E., Christopher, S., Lindstrom, R., Long, S., Marlow, A., Murphy, K., Paul, R., Popelka-Filcoff, R., Rabb, S., Sieber, J., 2010. Certification of three NIST renewal soil standard reference materials for element content: SRM 2709a San Joaquin Soil, SRM 2710a Montana Soil I, and SRM 2711a Montana Soil II. NIST Special Publication 260, 1-39.

Mahmood, H.S., Hoogmoed, W.B., van Henten, E.J., 2013. Proximal gamma-ray spectroscopy to predict soil properties using windows and full-spectrum analysis methods. Sensors 13, 16263-16280.

Manohar, S., Meijer, H., Herber, M., 2013. Radon flux maps for the Netherlands and Europe using terrestrial gamma radiation derived from soil radionuclides. Atmospheric environment 81, 399-412.

Mero, J.L., 1960. Uses of the gamma-ray spectrometer in mineral exploration. Geophysics 25, 1054-1076.

Peck, E., Bissell, V., 1973. Aerial measurement of snow water equivalent by terrestrial gamma radiation survey. Hydrological Sciences Journal 18, 47-62.

Pereira, L.S., 2011. Challenges on Water Resources Management when Searching for Sustainable Adaptation to Climate Change focusing Agriculture. European Water 34, 41-54.

Pracilio, G., Adams, M.L., Smettem, K.R., Harper, R.J., 2006. Determination of spatial distribution patterns of clay and plant available potassium contents in surface soils at the farm scale using high resolution gamma ray spectrometry. Plant and Soil 282, 67-82.

Priori, S., Bianconi, N., Fantappiè, M., Pellegrini, S., Ferrigno, G., Guaitoli, F., Costantini, E.A., 2013. The potential of γ-ray spectroscopy for soil proximal survey in clayey soils. EQA-International Journal of Environmental Quality 11, 29-38.

Söderström, M., Eriksson, J., 2013. Gamma-ray spectrometry and geological maps as tools for cadmium risk assessment in arable soils. Geoderma 192, 323-334.

Strati, V., Albéri, M., Anconelli, S., Baldoncini, M., Bittelli, M., Bottardi, C., Chiarelli, E., Fabbri, B., Guidi, V., Raptis, K.G.C., 2018. Modelling Soil Water Content in a Tomato Field: Proximal Gamma Ray Spectroscopy and Soil–Crop System Models. Agriculture 8, 60.

Sun, H., Nelson, M., Chen, F., Husch, J., 2009. Soil mineral structural water loss during loss on ignition analyses. Canadian journal of soil science 89, 603-610.

Szegvary, T., Leuenberger, M., Conen, F., 2007. Predicting terrestrial 222 Rn flux using gamma dose rate as a proxy. Atmospheric Chemistry and Physics 7, 2789-2795.

Van der Graaf, E., Limburg, J., Koomans, R., Tijs, M., 2011. Monte Carlo based calibration of scintillation detectors for laboratory and in situ gamma ray measurements. Journal of environmental radioactivity 102, 270-282.

Van der Klooster, E., Van Egmond, F., Sonneveld, M., 2011. Mapping soil clay contents in Dutch marine districts using gamma-ray spectrometry. European Journal of Soil Science 62, 743-753.

Viscarra Rossel, R., Taylor, H., McBratney, A., 2007. Multivariate calibration of hyperspectral γ-ray energy spectra for proximal soil sensing. European Journal of Soil Science 58, 343-353.

Vlastou, R., Ntziou, I.T., Kokkoris, M., Papadopoulos, C., Tsabaris, C., 2006. Monte Carlo simulation of γ-ray spectra from natural radionuclides recorded by a NaI detector in the marine environment. Applied Radiation and Isotopes 64, 116-123.

Walker, J.P., Willgoose, G.R., Kalma, J.D., 2004. In situ measurement of soil moisture: a comparison of techniques. Journal of Hydrology 293, 85-99.

Ward, S.H., 1981. Gamma-ray spectrometry in geologic mapping and uranium exploration. Economic Geology 75, 840-849.

Weynants, M., Montanarella, L., Toth, G., Arnoldussen, A., Anaya Romero, M., Bilas, G., Borresen, T., Cornelis, W., Daroussin, J., Gonçalves, M.D.C., 2013. European HYdropedological data inventory (EU-HYDI). EUR Scientific and Technical Research series.

Wilford, J., Bierwirth, P., e, Craig, M., 1997. Application of airborne gamma-ray spectrometry in soil/regolith mapping and applied geomorphology. AGSO Journal of Australian Geology and Geophysics 17, 201-216.

Wilford, J., Minty, B., 2006. The use of airborne gamma-ray imagery for mapping soils and understanding landscape processes. Developments in soil science 31, 207-610.

Xhixha, G., Bezzon, G., Broggini, C., Buso, G., Caciolli, A., Callegari, I., De Bianchi, S., Fiorentini, G., Guastaldi, E., Xhixha, M.K., 2013. The worldwide NORM production and a fully automated gamma-ray spectrometer for their characterization. Journal of Radioanalytical and Nuclear Chemistry 295, 445-457.

Zeng, W., Xu, C., Huang, J., Wu, J., Tuller, M., 2016. Predicting near-surface moisture content of saline soils from near-infrared reflectance spectra with a modified Gaussian model. Soil Science Society of America Journal 80, 1496-1506.